# Plackett-Burman experimental design for pulsed-DC-plasma deposition of DLC coatings

L. Pantoja-Suárez[1], M. Morales[2], J.L. Andújar[1], J. Esteve[1], M. Segarra[2], E. Bertran[1*]

[1] FEMAN Group, IN2UB, Departament de Física Aplicada i Óptica,
Universitat de Barcelona. C/ Martí i Franquès 1, 08028 – Barcelona (Spain)
[2] Centre DIOPMA, IN2UB, Departament de Ciència dels Materials i Enginyeria
Metal·lúrgica, Universitat de Barcelona. C/ Martí i Franquès 1, 08028 – Barcelona (Spain)

**ABSTRACT.** *The influence of technological parameters of pulsed-DC chemical vapour deposition on the deposition rate, the mechanical properties and the residual stress of diamond-like carbon (DLC) coatings deposited onto a martensitic steel substrate, using a Ti buffer layer between coating and substrate, has been studied. For this purpose, a Plackett-Burman experiment design and Pareto charts were used to identify the most significant process parameters, such as deposition time, methane flux, chamber pressure, power, pulse frequency, substrate roughness and thickness of titanium thin film. The substrate surfaces, which were previously cleaned by argon plasma, and the DLC coatings were characterized by scanning electron microscopy (SEM) and atomic force microscopy (AFM). The mechanical properties (elastic modulus and hardness) and the residual stress of DLC coatings were determined by the nanoindentation technique and calotte grinding method, respectively. Finally, the causes of the relative effect of different process variables were discussed.*

[*] Corresponding author:
E-mail address: ebertran@ub.edu (E. Bertran).

# 1. INTRODUCTION

Diamond-like carbon (DLC) is an amorphous material with $sp^2$ and $sp^3$ hybridized carbon atoms, which can also contain hydrogen.[1] The magnitude of the ratio of $sp^2$ and $sp^3$ bonds has a strong effect on both the mechanical and tribological properties of DLC.[2] Therefore, this carbon-based material is an excellent option to protect different kinds of surfaces. DLC coatings are widely used in different areas,[3] such as the injection moulding,[2] industries of automotive[4,5] and oil,[6,7] magnetic[8] and optical[9] applications, and medical applications.[10,11] The most representative mechanical properties required in protective materials are high hardness, elastic modulus hardness and good adhesion to the substrate.[12] However, the main drawback of DLC coatings is their high intrinsic compressive stress, to the order of several GPa, which alters their adhesion and limits the thickness of coatings.[13] Another problem is the difference in thermal expansion coefficients of DLC ($\sim 1\times10^{-6}$ $K^{-1}$) and steel ($\sim 11.8\times10^{-6}$ - $14.7\times10^{-6}$ $K^{-1}$). These two factors limit the maximum stable coating thickness that can be grown on steels. Previous studies have reported that the maximum DLC coating thickness is less than 0.25 µm on untreated steel surfaces.[14] When the DLC coating presents a poor adherence or the films are too thick, cracks may be formed, and the wear resistance is dramatically reduced, because the coating peels off.[2] Several alternatives have been studied to avoid the negative effect of internal stress. Doping with N, Si or metals, incorporating nanometric particles (DLC nanocomposites)[15,16] or post-annealing is one approach. Another is the deposition of alternated hard and soft a-C layers (a-C hard/a-C soft).[17,18] However, bias-graded deposition of DLC has the best tribological performance. This consists of surfaces showing a high $sp^3/sp^2$ ratio, which gradually decreases to the interface, in order to enhance their adhesion.[19] Another way consists in growing a buffer layer prior to DLC deposition. Bilayer systems help to reduce the stress of films and to enhance their scientific and industrial applications.[20] The most studied buffer layer materials are Ti, Cr, Cu, Ag, Ni and Si. The DLC/Ti system exhibits higher adhesion than other DLC/metallic systems, due to the formation of a defect-free interface, which is favourable energetically.[21] Another important feature of hydrogenated DLC coatings is their tribological behaviour, which has a low friction coefficient and low wear rate. The friction coefficient (against tungsten-steel) and wear rate against martensitic steel are in the range of 0.3-0.6 and $10^{-3}$ $mm^3$ $N^{-1}$ $m^{-1}$, respectively. In all environments, the tribological behaviour of DLC is controlled by an interfacial transfer layer formed during friction.[22] The build-up of a friction-induced transfer coating on the counterpart, followed by an easy shear within the interfacial material, is the most frequent friction-controlling mechanism for hydrogenated DLC coatings.[23]

Coating properties, such as hardness, wear resistance, residual stress and tribological behaviour, depend on both the method and the conditions of coating deposition. Additionally, tribological characteristics may likely be influenced by substrate roughness.[12] Therefore, the deposition process of DLC coatings represents a multi-variable system, which difficults the quantitative estimations of the effect of each parameter in the process. The relative effect of several variables on the coating properties needs to be investigated systematically, thus obtaining a quantitative estimation of the impact of each parameter on the process. When a great number of process parameters are involved, the statistics-based experimental designs are usually used in different fields related to biology, such as agriculture, biochemistry or medicine.[24,25] An economical and usual approach that provides information on the effects of single factors, but not on their interactions, is the Plackett-Burman (PB) method.[26,27] This method is well suited to ruggedness testing, thus establishing whether the outcome of an analytical procedure is affected by changes in each relevant factor. Therefore, it has been applied in many case studies of different industrial fields.[28,29,30] The most important feature of PB designs is that they involve $4n$ experiments, where $n$ = 1, 2, 3... In each case, the maximum number of factors that can be studied is $4n – 1$, so an 8-experiment PB design can study no more than 7 factors, a 12-experiment design will handle up to 11 factors, and so on. This may seem to be inconvenient, but it turns out to be a valuable feature of the method. Supposing four factors should be studied. Four experiments will be then insufficient. So, eight experiments in a PB design and seven factors shall have to use. This means that three of the latter will be dummy factors. They will have no chemical meaning at all. However it turns out that the apparent effects of these dummy factors can be used to estimate the random measurement errors. The more dummy factors there are, the better the estimate of such errors, so it is not uncommon for experimenters to use a larger PB design than is strictly necessary, thus getting higher quality information on the significance of each "real" factor. PB designs employ two levels for each factor, the higher level being denoted "+" and the lower "–" as usual. A further feature of the PB method is that the + and – signs for the individual trial experiments are assigned in a cyclical way. If eight experiments with seven factors labelled A–G are used, the levels for the first experiment might be: (A+ / B– / C+ / D– / E+ / F– / G+). Such sequences of + and – signs are provided by generating vectors, which are widely available in the literature and in software packages. The levels for the second experiment, again with four + and three – signs, are then obtained by moving the last sign for the first experiment to the beginning of the line, giving: (A+ / B– / C– / D– / E+ / F– / G+). This cyclical process is repeated for the first seven experiments. For the eighth experiment all the factors are set at the low (–) level, giving an overall design in which there are 28 + signs and 28 – signs, each factor having been studied four times at the higher level and four times at the lower. The effect of each factor is then readily determined from the expression (**1**):

$$\frac{2[\sum(y+)-\sum(y-)]}{N} \quad (1)$$

where *N* is the total number of experiments, eight in this case. The (*y*+) terms are the responses when a given factor is at its high level, and the (*y*-) terms reflect the responses for that factor set to its low level. Here, a proper choice of experimental conditions determines the success of the final evaluation. This may be shown through the experimental results. When DLCs are deposited, the diversity of the yield of the DLC coating should be considered an advantage, allowing correct assessment of the deposition operations used in the experiment.

The aim of the present study is to evaluate the influence of the technological parameters of DLC coating deposition that often affect the final properties of the coatings, such as mechanical properties (elastic modulus and hardness), residual stress, and deposition rate. DLC coatings were deposited by asymmetrical bipolar pulsed-DC chemical vapour deposition on the tuning surface of martensitic steel, using a Ti buffer layer between the DLC coating and the substrate. The mechanical properties and wear resistance of the DLC coatings were determined by the nanoindentation technique, using a continuous system measurement (CSM) module, and a calotte grinding method with a defined geometry, respectively. A Plackett-Burman experiment design was employed to preliminarily determine the effects of the typical process parameters, such as deposition time, methane flux, chamber pressure, power, pulse frequency, substrate roughness and thickness of titanium thin film. Although the deposition temperature strongly influences on the $sp^2/sp^3$ ratio and the hydrogen content, the effect of the deposition parameters has been analysed at room temperature in order to simplify the system.

## 2. Experimental procedure

### 2.1 Preparation of substrate surface and deposition process of DLC/Ti films

DLC/Ti coatings were deposited on disk substrates of a martensitic stainless steel grade AISI SAE M8 (high speed, molybdenum base) with a nominal hardness of 3 GPa.[31] **Table 1** shows the chemical composition of the steel. The initial substrates were polished to a surface roughness of $RMS = 3 \pm 2$ nm. Before deposition, the substrates were cleaned with ethanol in an ultrasonic bath for 10 min. The substrates were then ion-plasma cleaned in an argon atmosphere. In order to improve the adhesion of DLC coatings to the substrate, a 99.999% titanium target was used to grow a Ti buffer layer by magnetron sputtering. The parameters used to prepare the substrate surface and to deposit the titanium thin film are listed in **Table 2**.

**Table 1.** Chemical composition of AISI SAE M8 martensitic steel.

| Element | wt% |
|---------|-----|
| C | ≤ 0.8 |
| Si | ≤ 0.4 |
| Mn | ≤ 0.4 |
| Cr | 4.0-4.5 |
| Mo | 4.3-4.7 |
| W | 5.25-5.75 |
| Nb | 1.25 |
| V | 1.35-1.65 |
| Fe | Bal. |

After deposition of the Ti thin film and with a vacuum of $10^{-5}$ Pa prior to deposition of the DLC coating, methane was added, and the hydrogenated DLC top layer was deposited. The substrate was kept at room temperature by means of a water-cooling circuit, in order to grow the amorphous coatings. A set of experiments with Plackett-Burman (PB) experimental design was performed to develop the DLC coatings by the PECVD method. The PB designs involve a large number of factors and relatively few runs, and are typically used to identify a few significant factors out of a large set. A total of 12 experimental trials involving 7 independent variables were generated.[32] The deposition time, methane flux, chamber pressure, power, pulse frequency, substrate roughness (*RMS*) and thickness of titanium thin film interface were screened, as independent variables, within the ranges of 30-60 s, 20-30 sccm, 10-30 Pa, 50-120 W, 100-250 kHz, 2-6 nm and 20-100 nm, respectively. The details of these process parameters are listed in **Table 3**. The experimental data were used to plot Pareto charts, in order to identify the influence of independent variables on each response variable.

**Table 2.** Experimental conditions to prepare the substrate surface by Ar ion etching and Ti thin film deposition.

| Process | Time (s) | Ar flow (sccm) | Pressure (Pa) | Power (W) | Frequency (kHz) |
|---------|----------|----------------|---------------|-----------|-----------------|
| Ion etching | 900 | 30 | 5 | 50 | 250 |
|  | 1800 | 10 | 1 | 120 | 50 |
| Ti layer | 420 | 20 | 1 | 50 | 100 |
|  | 2040 | 20 | 1 | 50 | 100 |

**Table 3** Plackett-Burman experimental design matrix for screening the main parameters of DLC deposition.

| Code | Time (min) | Flux CH$_4$ (sccm) | Pressure (Pa) | Power (W) | Frequency (kHz) | Steel RMS roughness (nm) | Ti thickness (nm) |
|---|---|---|---|---|---|---|---|
| DLC-1 | 60 | 20 | 10 | 50 | 250 | 6 | 100 |
| DLC-2 | 60 | 30 | 30 | 50 | 250 | 6 | 20 |
| DLC-3 | 30 | 30 | 30 | 50 | 250 | 2 | 20 |
| DLC-4 | 30 | 30 | 30 | 120 | 100 | 6 | 100 |
| DLC-5 | 30 | 20 | 10 | 50 | 100 | 2 | 20 |
| DLC-6 | 30 | 30 | 10 | 50 | 100 | 6 | 100 |
| DLC-7 | 30 | 20 | 10 | 120 | 250 | 6 | 20 |
| DLC-8 | 60 | 20 | 30 | 120 | 100 | 6 | 20 |
| DLC-9 | 60 | 20 | 30 | 50 | 100 | 2 | 100 |
| DLC-10 | 30 | 20 | 30 | 120 | 250 | 2 | 100 |
| DLC-11 | 60 | 30 | 10 | 120 | 250 | 2 | 100 |
| DLC-12 | 60 | 30 | 10 | 120 | 100 | 2 | 20 |

Magnetron sputtering and plasma-enhanced CVD techniques, using pulsed-DC power for thin film growth, were employed to deposit thin film structures of DLC/Ti. Both processes were sequentially performed in one run, without breaking the vacuum. Sputtering targets were powered by a pulsed-DC signal in order to increase the ion energy of Ar, and hence the energy of Ti atoms and clusters emitted from the target. For the PECVD technique, the use of pulsed-DC power increases ion bombardment over the substrate during the positive pulse, which provides single energetic species and electrons in the plasma. The chemical reactions include the decomposition of methane, which is used as a carbon precursor in the experiments, to $CH_3$, $CH_2$, CH and C radicals, with $CH_3$ the most abundant carbon precursor.[1,33]

DLC and Ti films were deposited using PECVD technology and magnetron sputtering, respectively, powered with an asymmetrical bipolar pulsed-DC signal. The processing was carried out in a computer-controlled plasma reactor. The reactor chamber was pumped down by means of a turbomolecular pump, and the load-lock chambers were evacuated by rough pumping (up to ~5 Pa before opening the chamber valve). With this system, a base pressure of $10^{-5}$ Pa was achieved. The cathodes were connected to the pulsed-DC power supply to drive the voltage needed to switch-on and keep the plasma active. Gas valves, pressure gauges, and mass flow controllers were operated with a LabView interface, which can be programmed for each particular deposition process.[34]

## 2.2 Characterization of the substrate surface and DLC/Ti films

The morphological, mechanical and tribological properties of the surface substrate and DLC/Ti films were investigated using different techniques. The influence of the sputter-cleaning process on the nanostructure and morphology of the substrate surface was studied by scanning electron microscopy (SEM) (Jeol JSM-6510), equipped with an energy dispersive X-ray (EDX) detector. A Park XE-70 atomic force microscope (AFM) in non-contact mode was used to measure surface topography. The surfaces of the DLC coatings were also studied by AFM. In order to determine the value of the thickness of both the Ti and DLC films, profilometry was performed using a Dektak 3030. This technique mainly allows a study of surface roughness and thickness, due to its vertical resolution, in the order of 10 nm. The film thickness was estimated by measuring the height of a sharp step formed on the surface. Profilometer also measures the curvature of the substrate, which changes according to the stress of the DLC coating.

The residual stress of the films was determined by measuring the radius of curvature of the substrate before and after deposition using a surface profilometer, and by applying Stoney's equation:[35]

$$\sigma = \frac{E_S \cdot t_S^2}{6 \cdot (1-\upsilon_S) \cdot t_C} \left( \frac{1}{R} - \frac{1}{R_0} \right) \quad (2)$$

where $\sigma$ is the intrinsic mechanical stress; $E_S$ is Young's modulus of the substrate; $t_S$ is the substrate thickness; $\upsilon_S$ is Poisson's coefficient of the substrate; $t_C$ is the film thickness; $R$ is the curvature radius of the film/substrate system; and $R_0$ is the curvature radius of the substrate without film.

The mechanical properties of the films were characterized using the nanoindentation technique. It allows to obtain locally the mechanical properties, such as Young's modulus ($E$), hardness ($H$), fracture toughness ($K_{IC}$) and interfacial adhesion ($\sigma$), on small structural features in different types of materials as ceramics, metallic alloys, etc… and in bulk and film forms.[36,37,38,39,40] Nanoindentation tests were performed with a Nano Indenter® XP System (Agilent Technologies), equipped with Nanosuite 6.1 Explorer level software and a Continuous System Measurement (CSM) option that allows the continuous measurement of stiffness ($S$) and applied load ($P$) as a function of indentation depth ($h$). A three-sided pyramid Berkovich diamond indenter was used, which was calibrated with a fused silica standard (Young's modulus of 72 GPa).[41] The hardness and Young's modulus of DLC/Ti/martensitic steel as a function of indentation depth were obtained from the $P$-$H$ curves using the Oliver-Pharr algorithm.[42] A Poisson's ratio value of 0.30 for both the DLC film and the martensitic steel substrate was considered.[43] Indentations were performed up to a maximum indentation depth of 2000 nm and under a constant strain rate of 0.05 s$^{-1}$. More than fifty indentations were made in each material in four different regions, to achieve statistical significance. Subsequently, the $H$ of each film was directly determined in the range of indentation depths of around 10% of the film thickness[44,45,46] to avoid plastic interaction with the substrate, thus taking the maximum hardness as the one true value for each film. Young's modulus of each film was estimated by analysing $E$ at indentation depths below 10% of the film thickness, as the substrate usually influences the elastic response at lower indentation depths than the plastic one.[47] Therefore, the $E$ of the DLC films corresponded to the values of the plateaus obtained in this range of indentation depths.

Wear resistance was measured using Calotest Compact unit CSM Instruments 1. A steel ball of 30 mm diameter and under 60° angle of support was used to create grinding calottes with a defined geometry. A suspension of $Al_2O_3$ powder (1 µm) in glycerine was used as the grinding medium. Wear resistance measurements were compared to allow related statements within a series of samples applied to the same probing parameters.[13] All measurements were performed at 1.07 N and 320 rpm for 120 s. Archard's equation,[48] as shown in (**3**), provides the wear rate, $w$, from the used input parameters and the depth of the crater:

$$w = \frac{V}{s \cdot F} = \frac{1/3 \cdot \pi \cdot h^2 \cdot (3R - h)}{\omega \cdot R \cdot t \cdot F} \qquad (3)$$

where $V$ is the crater volume, $s$ is the length of the path travelled by the ball; $F$, the normal force applied to the surface; $h$, the depth of the crater; $R$, the ball radius; $\omega$, the angular velocity of the ball; and $t$, the wear time.

## 3. Results and discussion

### 3.1 Effect of Ar ion-plasma cleaning on substrate surface roughness

After exposure of the substrate to cleaning by argon ion-plasma, images of SEM and AFM revealed morphological and nanostructural changes in the substrate surface. **Fig.1a** shows an SEM image of the substrate obtained by the secondary electron detector. While the bright zone corresponds to the metallographically polished steel region, the dark one corresponds to the Ar ion-plasma treated region. In **Fig.1b,** the SEM micrograph, obtained by the backscattered electron detector, clearly evidences the presence of precipitates with a size up to 15 µm in the matrix phase. **Figs.1c** and **d** present the energy dispersive

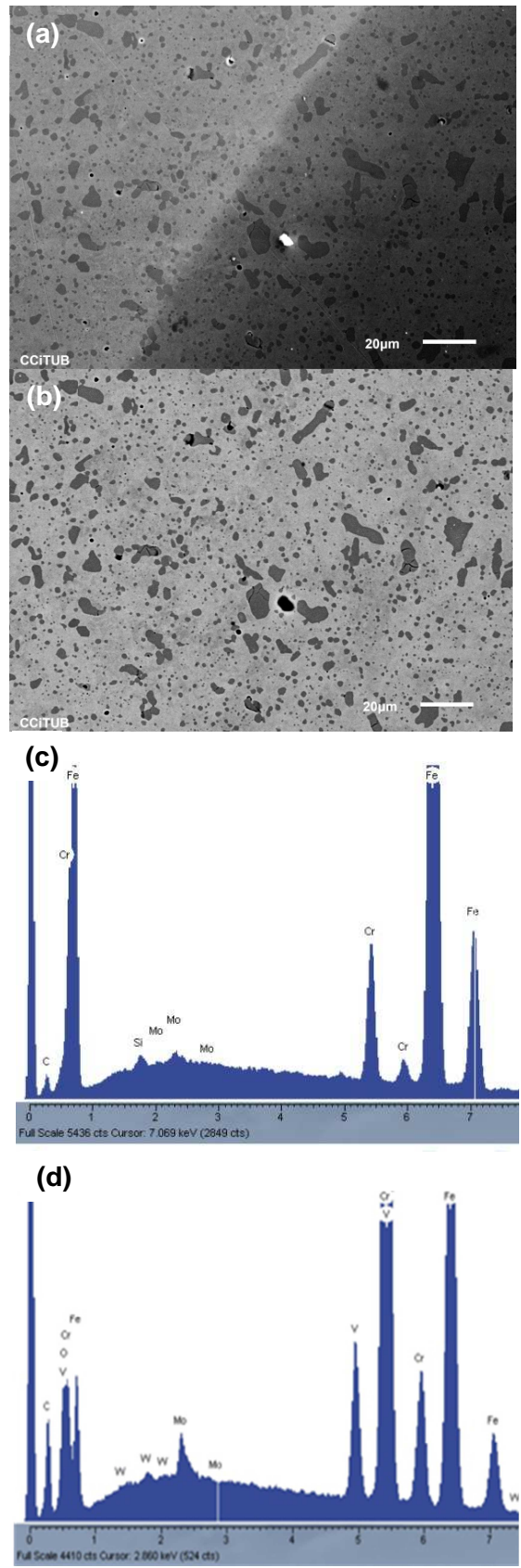

**Fig.1** SEM images of the polished substrate surface obtained by: (a) secondary electron detector, and (b) backscattered electron detector. EDX analysis for: (c) steel matrix, and (d) precipitates.

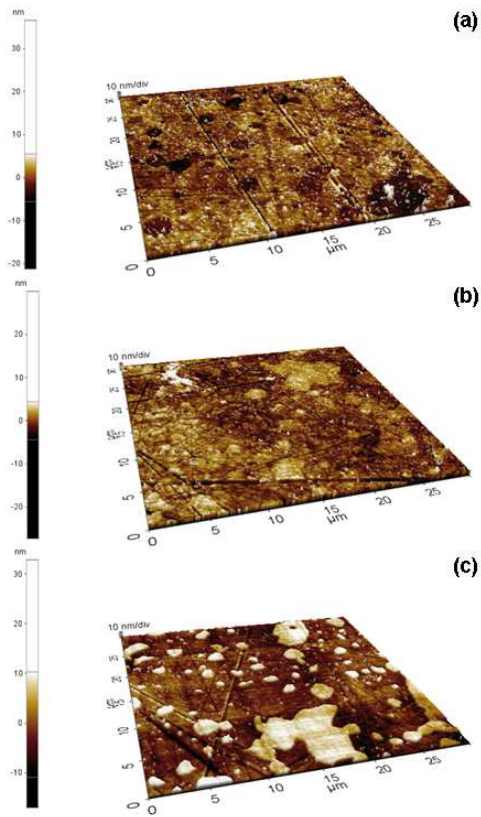

**Fig.2** AFM images of substrate roughness: a) before argon plasma etching; and after argon plasma etching with: b) RMS = 2±1 nm, and c) RMS = 6±1 nm.

X-ray (EDX) spectra for the matrix and the precipitate phases, respectively. The main difference between both phases is the higher content in carbon, chromium, tungsten and vanadium at the precipitates, thus forming the typical carbides of these metals. To obtain a smooth and rough substrate surface, preliminary experiments were performed to determine the ion-plasma etching technological parameters (**Table 2**).

**Fig.2** shows AFM images of the substrate surface before (**Fig.2a**) and after the argon ion-plasma cleaning process (**Fig.2b** and **c**). Smooth surfaces were obtained with roughness values (expressed in *RMS*) of 2 nm (**Fig.2b**) and 6 nm (**Fig.2c**). All images exhibit several metallographic polish marks. However, the images after argon etching show that the matrix phase was more clearly etched than the precipitates. Therefore, the results indicate that precipitates with an appreciable content of V and Cr are harder than the matrix phase. Consequently, ion bombardment induces selective etching of the substrate, which generates surface enrichment of V and Cr and surface roughness that replicates the domains of precipitates.

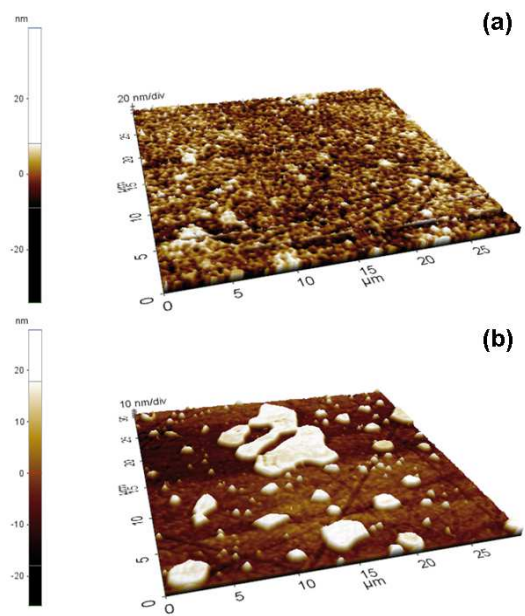

**Fig.3** AFM images of DLC films roughness: a) sample DLC-11 with RMS = 4±1 nm, and b) sample DLC-8 with RMS = 9±1 nm.

### 3.2 Surface roughness of DLC films

Thin films of titanium with 20 and 100 nm thicknesses were deposited onto the previously polished martensitic steel surfaces. DLC films with thicknesses over 0.4 µm were subsequently deposited onto Ti thin films. AFM images of two samples of DLC films with a smooth surface (substrate roughness ~ 2 nm) and a rough surface (substrate roughness ~ 6 nm) are shown in **Fig.3a** and **b**, respectively. Although the thickness of the DLC film was greater than 0.4 µm, these results indicate that the surface roughness of DLC depends on the surface roughness of the substrate.

### 3.3 Deposition rate of DLC

In **Fig.4**, the Pareto charts indicate the standardized effect of process parameters on the deposition rate, the mechanical properties and the residual stress of DLC coatings. The length and colour of the bars correspond to the intensity grade, and the positive (grey) or negative (blue) effect of the process parameters. As shown in **Fig.4a**, the Pareto chart reveals that an increase in both the power and the pressure of the PECVD process had a positive, statistically significant effect on the deposition rate of DLC coatings. In contrast, substrate roughness did not have a remarkable effect on the deposition rate. The strong effect of power can be attributed to the fact that the ionisation energies and rates achieved by pulsed-DC methane have a direct impact on electron density, thus resulting in more chemically active plasma.[19] The system presented a pulsed regime for generating ions with the time, thus producing strong accumulations of charge with increasing the ionization level, and so activating the break activity of species. If the precursor is optimized, it influences on the increase of the plasma density, which is correlated with the density of species that produces the growth of layer.[49,50] **Table 4** shows the deposition rates of DLC coatings for all performed experiments. The highest rate corresponds to the DLC-8 sample, which was obtained with a deposition rate of 54 nm/min and with a thickness of around 3200 nm. In contrast, the DLC-1 sample was deposited at a lower rate close to 11 nm/min and has a thickness of around 660 nm. The study of the microscopic growth mechanisms and hence the structural properties of the films are not an objective of this work, because a complete study about the dependency of the deposition rate with the frequency of pulsed-DC plasma was already presented in other work for similar systems as reported by Corbella et al.[1] According to this work, the high frequencies produce deposition rates lower than those of radiofrequency technology (RF-PECVD). It can be also observed that the low frequencies of pulsed-DC plasma technology correspond to the highest deposition rates. For a power around 100 W, the deposition rates were of 23 and 52 mm/min at frequencies of 200 kHz and 100 kHz, respectively. In addition, traditional explanation for the presence of the large internal stress is based upon the assumed sub-implantation mechanism of DLC deposition.[51] In this mechanism, the high energetic incident species of carbon bombard the surface of DLC film at room temperature. Consequently, the low mobility of carbon atoms of the DLC film results in a high compressive stress in the layer.

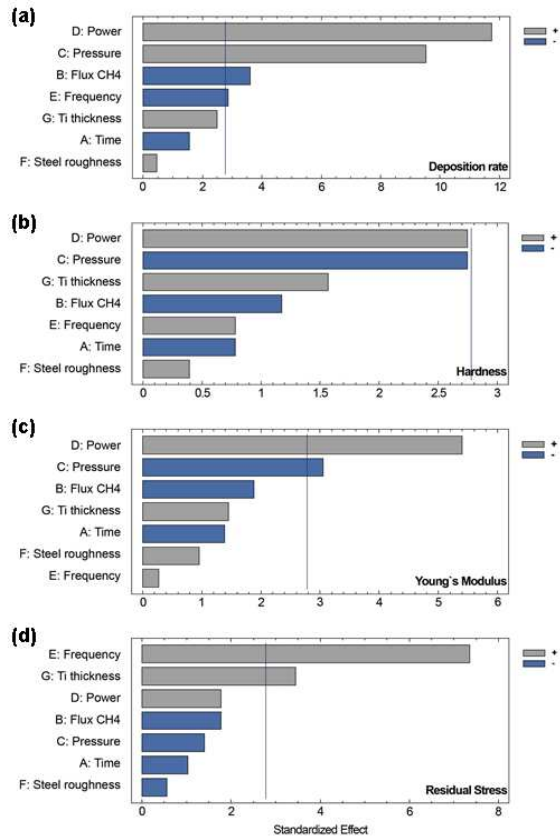

**Fig.4** Pareto chart of the standardized effect of independent variable for (a) deposition rate, (b) hardness, (c) Young´s modulus and (d) residual stress of DLC films.

**Table 4** Summary of different properties obtained for the hydrogenated DLC films.

| Code | Deposition rate (nm/min) | DLC thickness (nm) | Stress (GPa) | Hardness (GPa) | Young`s Modulus (GPa) |
| --- | --- | --- | --- | --- | --- |
| DLC-1 | 11 | 668 | 0.48 | 18.3 ± 1.2 | 126 ± 11 |
| DLC-2 | 18 | 1061 | 0.34 | 11.9 ± 0.9 | 83 ± 7 |
| DLC-3 | 22 | 662 | 0.27 | 12.8 ± 0.9 | 86 ± 7 |
| DLC-4 | 52 | 1561 | 0.21 | 15.9 ± 1.2 | 124 ± 10 |
| DLC-5 | 16 | 473 | 0.21 | 16.3 ± 0.9 | 117 ± 7 |
| DLC-6 | 15 | 438 | 0.25 | 15.7 ± 1.4 | 112 ± 11 |
| DLC-7 | 33 | 995 | 0.47 | 17.0 ± 0.5 | 124 ± 5 |
| DLC-8 | 54 | 3240 | 0.13 | 16.3 ± 0.4 | 132 ± 6 |
| DLC-9 | 35 | 2116 | 0.20 | 11.7 ± 0.9 | 88 ± 8 |
| DLC-10 | 51 | 1523 | 0.64 | 18.6 ± 1.3 | 137 ± 9 |
| DLC-11 | 30 | 1789 | 0.53 | 17.8 ± 0.7 | 128 ± 6 |
| DLC-12 | 24 | 1442 | 0.15 | 16.0 ± 0.6 | 122 ± 6 |

### 3.4 Hardness and Young`s Modulus

**Fig.5a** shows the evolution of the hardness ($H$) of the DLC/Ti/martensitic steel system as a function of the normalized indentation depth, in terms of indentation depth/coating thickness ($h/t$). Three different zones could be observed in all samples, as typically obtained for a hard coating on a soft substrate. At small normalized indentation depths, hardness was strongly increased. Subsequently, it reached a maximum value in the $h/t$ range between 0.03 and 0.2, in which the plastic response is only attributed to the coating properties. Depending on the coating thickness and its behaviour at the interface in contact with the Ti interlayer and substrate, $H$ tended to stabilize to a constant value in the normalized indentation depth range of 0.1 to 0.2. As the normalized indentation depth increased, $H$ progressively decreased, which is associated with the mixed plastic response of the DLC coating, Ti interlayer and the substrate. Finally, the plastic response was mainly dominated by the substrate at higher normalized indentation depths.

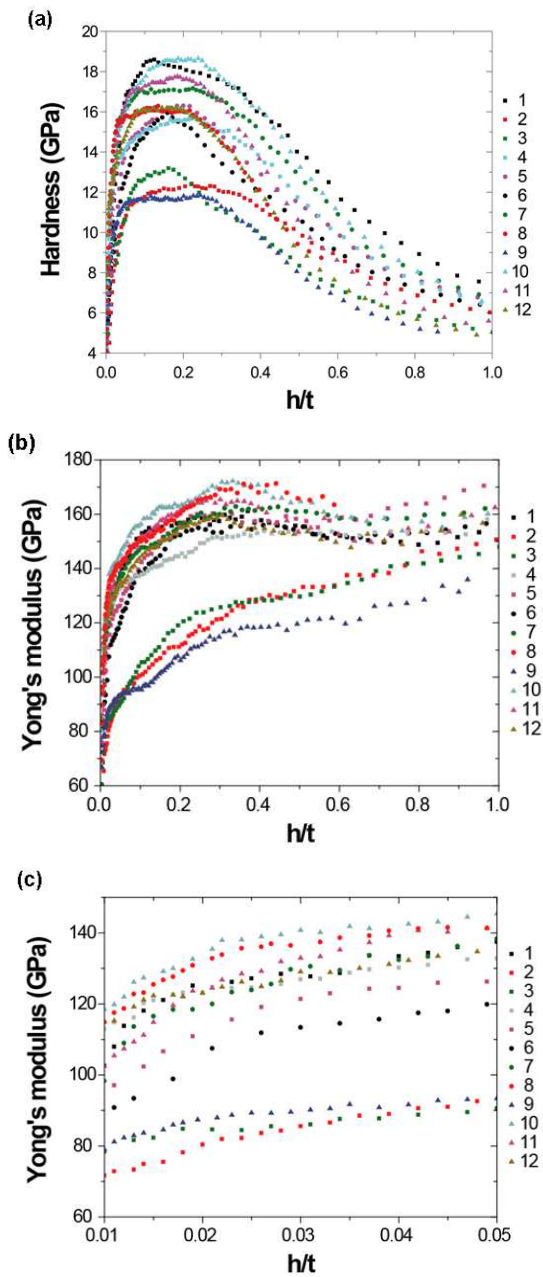

**Fig.5** a) Hardness as a function of normalized indentation depth for DLC/Ti/martensitic steel systems; and Young's modulus as a function of normalized indentation depth in the range off : b) $0 < h/t < 1$ and c) $0 < h/t < 0.05$, for DLC/Ti/martensitic steel systems.

**Fig.5b** shows the evolution of Young's modulus in the DLC/Ti/substrate system, which depends on both Young's modulus of the coating, Ti interlayer and substrate, and the behaviour of the coating interface in contact with the Ti interlayer and substrate. When $E$ of both the Ti interlayer ($\sim$ 120 GPa)[52] and substrate ($\sim$ 200 GPa determined by our preliminary experiments) were above that of the DLC coating, the apparent Young's modulus gradually rose with the increase in normalized indentation depth. In contrast, $E$ increased to stabilize at an almost constant value for $h/t > 0.1$-$0.3$ for DLC coatings with a Young's modulus close to the Ti interlayer and lower than that of the steel substrate. **Fig.5c** confirms that small indentation depths are required to determine and isolate the Young's modulus of the DLC coatings correctly. This can be attributed to that the substrate influences on the elastic response at lower indentation depths than the plastic one, as the elastic field has a larger volume than that produced by plastic deformation. From the analysis, all curves of Young's modulus versus normalized indentation depth present plateaus between 0.01 - 0.02 and 0.05. In the initial stage ($h/t < 0.0 - 0.02$), this is influenced by a number of effects, such as tip rounding, indenter drift, determination of the zero point, surface roughness and machine resolution. Therefore, we considered the plateau values as the true Young's modulus of the DLC coatings, since the other stages belong to transitional regions including the coating, Ti interlayer and substrate characteristics. **Table 4** lists the $E$ and $H$ values of the different coatings. Both the mechanical properties and the adhesion tests evidence that the DLC-8 sample with a thickness of 3200 nm exhibits the best mechanical behaviour of the studied samples. The low dispersion of both mechanical properties obtained in different zones reveals the high homogeneity and quality of this coating. DLC-8 presents a hardness of $16.3 \pm 0.4$ GPa and a Young's modulus of $132 \pm 6$ GPa. Consequently, a relatively low $E/H$ ratio (around 8) is achieved, as suggested by the significant elastic recovery observed in the typical load-depth curve. In addition, some pop-in events at high loads on the sample are observed. This may represent the formation and propagation of radial cracks, due to tensile stress concentrations at the coating/substrate interface during the load process, which have been identified as a driving force behind this type of crack.[38,53] The mechanical properties of DLC coatings (**Table 4**) are comparable to those of literature reported in similar deposition conditions and systems.[54,55,56] Choi et al.[54] determined hardness values ranged of 10 – 16 GPa and Young's modulus of 130 - 155 GPa for DLC coatings prepared by a RF-PECVD method, at room temperature, on glass and silicon substrates by using methane and hydrogen gas with different deposition conditions. Bec et al.[55] presented a hardness of $10.6 \pm 1$ GPa and a reduced Young's modulus of $80 \pm 5$ GPa for a DLC coating obtained by pulsed-DC PECVD at room temperature using acetylene as a precursor gas. Pancielejko et al.[56] obtained DLC coatings, depending from the used deposition parameters, with a wide hardness range of 20 – 60 GPa at temperatures from RT to 200ºC.

**Fig.4b** shows that, in the present study, the power (positive effect) and the pressure (negative effect) mainly influenced the hardness of the DLC coatings. Both independent variables just reached statistical significance. However, these significances were much higher than those for the other analysed parameters. Hence, both power and pressure were considered the main parameters controlling the hardness of the DLC coatings. The next Pareto chart, **Fig.4c**, also evidences the strong influence (positive effect) of the power plasma excitation on the value of Young`s modulus for DLC coatings. This effect can be attributed to the increase in ion bombardment and ion energy during the DLC growing process. As commented in section 3.3, our system uses a non-stationary regime, which is a pulsed regime for generating and neutralizing ions with the time. The optimization of precursor produces strong accumulations of charge, thus increasing the ionization level and therefore the density of species that produces the growth of layer.[49,50]

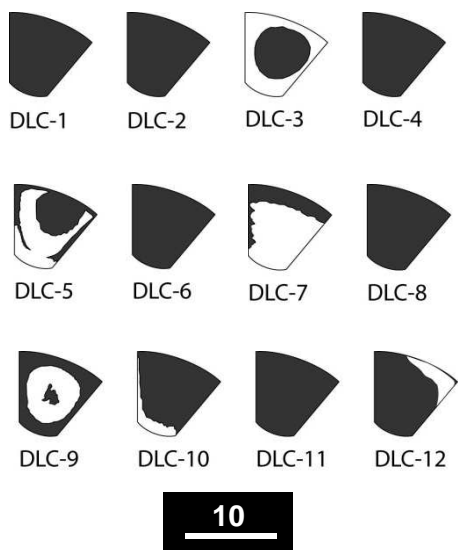

**Fig.6** Illustration of DLC/Ti films on martensitic steel, as deposited. After DLC peeling off, black and white areas correspond to DLC/Ti film and martensitic steel, respectively.

### 3.5 Residual stress and peeling off

Residual stress in the DLC coatings varies in the range of 0.13 - 0.64 GPa. These positive values correspond to compressive stress. Both the hardness and compressive stress values are a consequence of ion bombardment during DLC growth. The application of pulsed-DC technology has several advantages over other technologies such as RF-PECVD.[1] For example, higher power density and control of the frequency and duty cycle of the DC pulses are achieved. The sample can relax during the time of the negative pulse, which reduces the stress before the next pulse is received. In **Fig.4d,** the Pareto chart confirms that the frequency of the pulsed-DC signal has the highest positive effect on the stress reduction of the DLC coatings. In addition, the Ti thin film leads to considerable improvements in the adherence of the stressed DLC coatings, because it acts as a buffer layer, reducing the density of mechanical energy accumulated in the DLC coatings. The duty cycle of the DC power supply was kept at around 20% for all the experiments. Low stress values can be obtained using suitable time intervals in the positive and negative pulses 1. In our case, low stress was achieved when the time of the negative pulses increased from 3.2 µs to 16 µs.

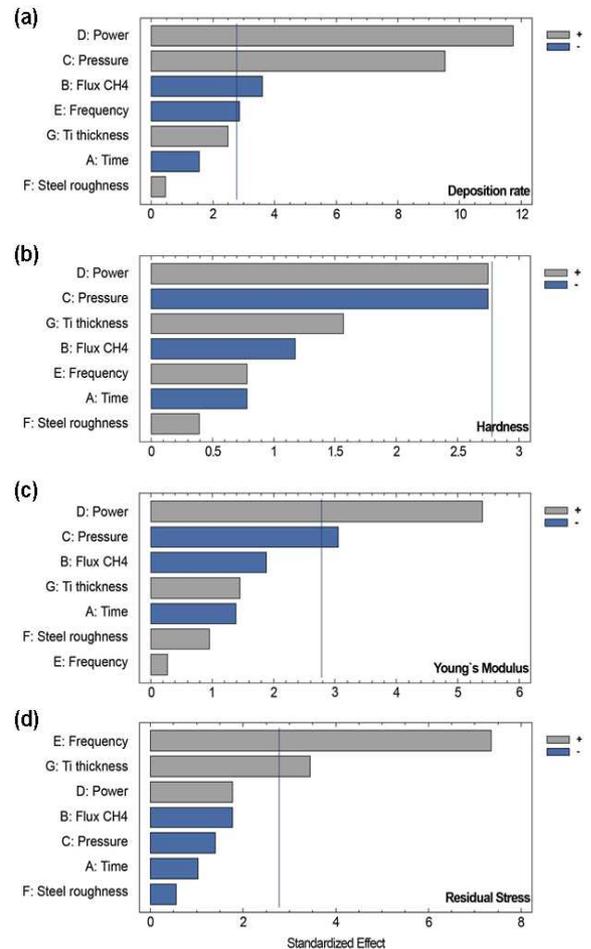

**Fig.7** Pareto chart of the standardized effect for adherence.

**Fig.6** illustrates the deposited material for each experiment. Seven bilayers appear to be well-adhered to the steel substrate at different conditions. Black and white areas correspond to the DLC/Ti film and martensitic steel, respectively. However, in the rest of the samples, the DLC/Ti bilayer was peeled off from the substrate. **Fig.7** shows the Pareto chart related to the influence of the primary parameters on the adherence between the DLC/Ti bilayer and martensitic steel. In this analysis, two numerical values were considered to generate the Pareto chart. Samples with total adherence were assigned with number 1 and samples with some peeling off were associated with number 0. In this analysis, the surface roughness was found to be the only statistically significant parameter. As mentioned in the introduction, high values of residual stress do not allow DLC coating growth over 0.25 µm.[14] In this study, a stable DLC coating was obtained with around 3 µm of thickness and good mechanical properties, as in the DLC-8 sample, which had the lowest residual stress value (0.13 GPa). In summary, the residual stress of DLC coatings was reduced from 2 – 3 GPa for a standard DLC to below 1 GPa by controlling the frequency of the pulsed-DC signal and the thickness of the Ti buffer layer. For these small stress values, the roughness of the martensitic steel appears to be the most critical parameter to control the adhesion.

### 3.6 Wear resistance

Wear resistance was measured in eight DLC/Ti bilayer samples (**Fig.8**) in which the substrate area was almost or completely coated. The wear rate was evaluated from the geometry of the calotte of the DLC/Ti bilayers, using **Eq. (2)**. Coatings with Ti interlayers of 100 nm thickness, except for the DLC-9 sample, presented wear rates within the range of $10^{-4}$ - $10^{-8}$ mm$^3 \cdot$N$^{-1}$m$^{-1}$. The DLC-4 sample had the lowest wear rate ($3.7 \times 10^{-8}$ mm$^3$ N$^{-1}$m$^{-1}$), as it presented substrate surface roughness and Ti film thickness of 6 and 100 nm, respectively. Therefore, high values of both surface roughness and Ti buffer layer strongly decrease the wear rate of the DLC coatings. Although the DLC-8 sample had a Ti interlayer thickness of 20 nm, it also presented a low wear rate ($1.17 \times 10^{-7}$ mm$^3$N$^{-1}$m$^{-1}$). This may be due to the effect of high substrate roughness and low residual stress. These wear rates are comparable to those reported by other authors obtained by similar technologies.[56,57] Pancielejko et al.[56] showed DLC coatings deposited on an HS6-5-2 steel substrates, using pulsed-DC PECVD, which presented a wear rate range of 1.2 - 6.9 $\times 10^{-7}$ mm$^3$ N$^{-1}$m$^{-1}$ at deposition temperatures from room ones to 200ºC, respectively. Xiang et al. [57] obtained wear rates in a range of 2.8 - 8.7 $\times 10^{-8}$ mm$^3$N$^{-1}$m$^{-1}$ for DLC films deposited in mid-frequency dual-magnetron sputtering.

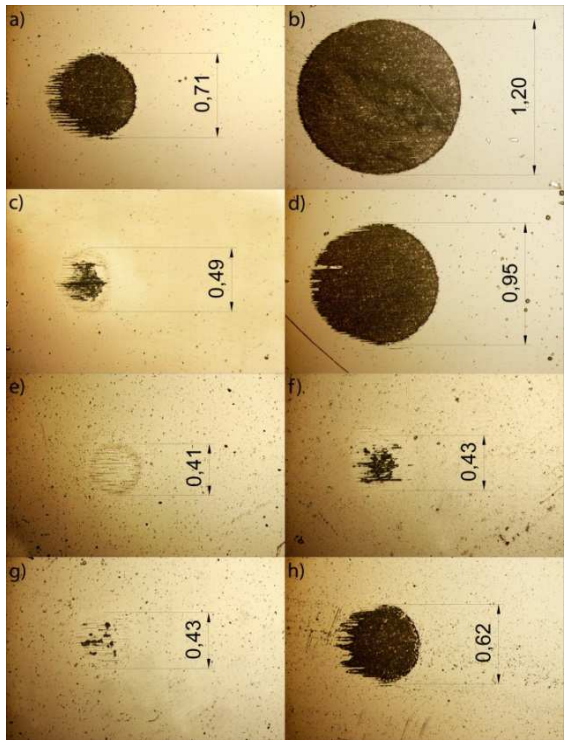

**Fig.8** Optical micrographs of wear scars formed on the DLC films after the calotest: a) DLC-1, b) DLC-2, c) DLC-4, d) DLC-6, e) DLC-8, f) DLC-10, g) DLC-11, and h) DLC-12 (units in mm).

## 4 Conclusions

A Plackett-Burman experimental design and Pareto charts were used to analyse the pulsed-DC PECVD technological parameters that influence on the deposition rate, the mechanical properties and the wear resistance of DLC coatings deposited on a Ti film/martensitic steel system. This was found to be a suitable and useful approach. A set of experiments based on a simple experimental design enabled us to identify the effect of the main process variables involved in the DLC coating deposition by pulsed-DC PECVD. The results indicate that pulsed-DC PECVD power, depositing the DLC coating at room temperature, is the main technological parameter that influences the deposition rate, the hardness and Young`s modulus of the DLC coatings. The deposition rate and the mechanical properties rose significantly with an increase in pulsed-DC PECVD power. However, pulse frequency was the most critical parameter to decrease residual stress. Low intrinsic residual stress was achieved by controlling the frequency and duty cycle of the DC pulses. In addition, when the DLC coating had a low stress value, the surface roughness of the steel substrate was the main parameter that strongly influenced adhesion and residual stress, which is related to wear resistance, in the DLC/Ti/steel system. Finally, the mechanical properties and wear rates of these DLC coatings were comparable to those of literature reported in similar technologies and deposition conditions.

**Acknowledgments**

This work was financed by project 2014SGR01543 and 2014SGR0948 from AGAUR of the Generalitat de Catalunya, MAT2010-20468 from MICINN of the Spanish Government, project UNBA10-4E-316 financed by FEDER-EU, and SENESCYT of the Ecuadorian Government, which provided financial support through its scholarship program for 2011. The authors thank Mrs. Toffa Evans of the Language Services of the University of Barcelona for language revision.